\documentclass[letterpaper,twocolumn,english,aps,prl,preprintnumbers,groupaddress,twocolumn,nofootinbib]{revtex4}
\usepackage[T1]{fontenc}
\usepackage[latin9]{inputenc}
\usepackage{float}
\usepackage{amsmath}
\usepackage{graphicx}

\makeatletter

\newcommand{\noun}[1]{\textsc{#1}}

\@ifundefined{textcolor}{}
{%
 \definecolor{BLACK}{gray}{0}
 \definecolor{WHITE}{gray}{1}
 \definecolor{RED}{rgb}{1,0,0}
 \definecolor{GREEN}{rgb}{0,1,0}
 \definecolor{BLUE}{rgb}{0,0,1}
 \definecolor{CYAN}{cmyk}{1,0,0,0}
 \definecolor{MAGENTA}{cmyk}{0,1,0,0}
 \definecolor{YELLOW}{cmyk}{0,0,1,0}
 }

\@ifundefined{definecolor}{\@ifundefined{definecolor}{\@ifundefined{definecolor}{\@ifundefined{definecolor}{\@ifundefined{definecolor}
 {\@ifundefined{definecolor}
 {\usepackage{color}}{}
}{}
}{}}{}}{}}{}

\makeatother\usepackage{babel}\makeatother\usepackage{babel}\usepackage{algorithm}\usepackage{algpseudocode}

\makeatother

\usepackage{babel}

\makeatother

\usepackage{babel}

\makeatother

\usepackage{babel}

\makeatother

\usepackage{babel}

\begin{document}

\title{Fast Multiplication of Matrices with Decay}

\author{Matt Challacombe and Nicolas Bock}

\email{{mchalla,nbock}@lanl.gov}

\affiliation{Theoretical Division, Los Alamos National Laboratory, Los Alamos,
New Mexico 87545}

\homepage{freeon.org}
\begin{abstract}
A fast algorithm for the approximate multiplication of matrices with
decay is introduced; the Sparse Approximate Matrix Multiply (SpAMM)
{\normalsize reduces complexity in the product space,} a different
approach from current methods that economize within the matrix space
through truncation or rank reduction.{\normalsize{} Matrix truncation
(element dropping) is compared to SpAMM for quantum chemical matrices
with approximate exponential and algebraic decay. For matched errors
in the electronic total energy, SpAMM is found to require fewer to
far fewer floating point operations relative to dropping. The challenges
and opportunities afforded by this new approach are discussed, including
the potential for high performance implementations. }{\normalsize \par}
\end{abstract}

\preprint{\texttt{LA-UR} 10-07458}

\keywords{Fast Matrix Multiplication, Electronic Structure, Matrix Function,
Matrix Decay, Spectral Projection, Generalized N-Body Problem, Sparse
Matrix-Matrix Multiply, Linear Scaling Quantum Chemistry}

\maketitle

\section{Introduction}

For large dense linear algebra problems, the computational advantage
offered by fast matrix-matrix multiplication can be substantial, even
with seemingly small gains in asymptotic complexity. Relative to conventional
multiplication which is $\mathcal{O}\left(n^{3}\right)$, Strassen's
algorithm achieves $\mathcal{O}\left(n^{2.8}\right)$, while the Coppersmith
and Winograd method is $\mathcal{O}\left(n^{2.38}\right)$. For these
dense methods, balancing the trade off between cost, complexity and
error is an active area of research \cite{Demmel:1992:FastMM,Demmel:2007:FastMM,Yuster:2005:FastMM}.

On the other hand, large sparse problems are typically handled with
conventional sparse matrix techniques, with only small concessions
between multiplication algorithms. Intermediate to these regimes,
a wide class of problems exist that involve matrices with decay%
\footnote{A matrix $A$ is said to decay when its matrix elements decrease exponentially,
as $|a_{i,j}|<c\lambda^{|i-j|}$ , or algebraically as $|a_{i,j}|<\frac{c}{|i-j|^{\lambda}+1}$
with indicial separation $|i-j|$. In non-synthetic cases, the separation
$|i-j|$ typically corresponds to an underlying physical distance
$|\vec{r}_{i}-\vec{r}_{j}|$, \emph{e.g.} of basis functions, finite
elements, \emph{etc}. See Figure \ref{fig:Decay-of-normed} as well
as the excellent work by Benzi and co-authors on this topic in References
\cite{Benzi:2007:Decay,Benzi:2010:Decay}. %
}where sparsity exists only asymptotically under an approximate linear
algebra, historically involving matrix economization through element
dropping or rank reduction. Often, problems with decay occur in the
construction of matrix functions, notably the matrix inverse \cite{Benzi:2000:Inv},
the matrix exponential \cite{Iserles:1999:Decay}, and in the case
of electronic structure theory, the Heaviside step function \cite{Challacombe:1999:DMM,Challacombe:2000:SpMM,Benzi:2007:Decay,Benzi:2010:Decay}.
The use of an approximate matrix algebra is also an active area of
interest in the solution of large eigenproblems \cite{Simoncini:2002:Inexact,Simoncini:2003:Inexact,Simoncini:2005:Inexact,Challacombe:2010:TDSCF}.

Many approaches to a sparse approximate linear algebra exist for matrices
with decay \cite{Galli:1996:ONREV,Goedecker:1999:ONREV,Goedecker:2003:ONREV,Li:2005:ONREV},
largely predicated upon the truncation of matrix elements, with the
recent work of Benzi providing the most detailed analysis so far \cite{Benzi:2007:Decay,Benzi:2010:Decay}.
In this contribution, we develop sparse matrix multiplication as a
generalized $N$-body problem \cite{Gray:2001:NBody}, and introduce
a fast algorithm based on hierarchical truncation in the three-dimensional
space $i,\, j,\, k\in[1,n]$ of the product $C_{ij}=\sum_{k}A_{ik}B_{kj}$,
where $A$ and $B$ decay exponentially or algebraically fast enough%
\footnote{Algebraic decay sufficient to achieve a fast $\mathcal{O}\left(n\,\lg\, n\right)$
or better complexity is an open question similar to that of conditional
convergence in the shape dependent summation of dipole and quadrupole
components of the Lorentz field. %
}. Viewing the product from a length scale perspective$^{1}$, if matrix
elements decay as $\mathcal{O}(1/r^{\lambda})$, then the bulk of
the product interactions will decay as $\mathcal{O}(1/r^{2\lambda})$.
For small $\lambda$, the difference between truncation in the matrix
space and the product space may be significant.

\section{A Sparse Approximate Matrix Multiply}

\begin{figure}[H]
\includegraphics[width=3.3in]{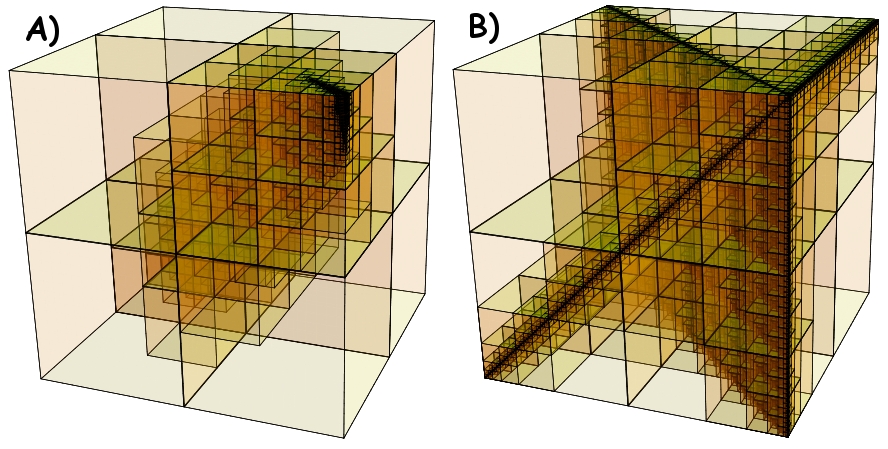}\caption{Hierarchical truncation of the product space ($i,\, j,\, k$) using
$\tau=10^{-8}$ for synthetic matrices of dimension $n=512$, with
$A_{ij}=\exp\left(-\left|i-j\right|\right)$ and $B_{ij}=\exp\left(-2\left|i-j\right|\right)$
in \textsf{\textbf{(A)}}, $A_{ij}=B_{ij}=\protect\begin{cases}
1/|i-j|^{3} & i\neq j\protect\\
0 & \textrm{else}\protect\end{cases}$ in \textsf{\textbf{(B)}}. Each box above the finest ($k=k_{max}$)
scale represents truncation. \label{fig:Hierarchical-truncation-of}}

\end{figure}

The quadtree matrix representation, 

\begin{equation}
A^{k}=\left(\begin{array}{cc}
A_{11}^{k+1} & A_{12}^{k+1}\\
A_{21}^{k+1} & A_{22}^{k+1}\end{array}\right),\: k=0,\,\ldots,\, k_{{\rm \textrm{max}}},\end{equation}
 is the basis for recursive matrix-matrix multiplication, $C^{k}=A^{k}\cdot B^{k}.$
For conventional recursive multiplication, the operator ``$\cdot$''
is just the row-column product, while in fast multiplication, it represents
an economized sequence of operations with reduced complexity and a
more complicated error accumulation. In Reference \cite{Bini:1980:FastMM},
Bini and Lotti carried out a detailed error analysis for recursive
matrix multiplication schemes, and derived component-wise bounds of
the form

\begin{equation}
|\tilde{c}_{ij}-c_{ij}|<a\, b\,\epsilon\, n\,\lg_{2}\, n\end{equation}
where $\tilde{c_{ij}}$ is a matrix element computed to within precision
$\epsilon$, $c_{ij}$ is its exact counterpart, $a=\max_{ij}|a_{ij}|$
and $b=\max_{ij}|b_{ij}|$. While providing a sharp bound, the max
norm does not immediately lend itself to the recursive separation
of interaction magnitudes. Consider instead the framework provided
by an arbitrary sub-multiplicative matrix norm $\Vert\cdot\Vert$:

\begin{eqnarray}
\Vert C^{k}\Vert & \leq & \Vert A^{k}\Vert\Vert B^{k}\Vert\nonumber \\
 & \leq & \Vert A_{11}^{k+1}\Vert\Vert B_{11}^{k+1}\Vert+\Vert A_{12}^{k+1}\Vert\Vert B_{21}^{k+1}\Vert\label{eq:SubMultiplicative}\\
 & + & \Vert A_{11}^{k+1}\Vert\Vert B_{12}^{k+1}\Vert+\Vert A_{12}^{k+1}\Vert\Vert A_{22}^{k+1}\Vert\nonumber \\
 & + & \Vert A_{21}^{k+1}\Vert\Vert B_{11}^{k+1}\Vert+\Vert A_{22}^{k+1}\Vert\Vert B_{21}^{k+1}\Vert\nonumber \\
 & + & \Vert A_{21}^{k+1}\Vert\Vert B_{12}^{k+1}\Vert+\Vert A_{22}^{k+1}\Vert\Vert A_{22}^{k+1}\Vert\,.\nonumber \end{eqnarray}
This structure suggests an algorithm we call the Sparse Approximate
Matrix Multiply (SpAMM), which recursively tests each of the 8 contributions
in Equation~(\ref{eq:SubMultiplicative}) for significance in accordance
with a given numerical threshold $\tau$:

\begin{widetext} \begin{equation}
{\tt SpAMM}(A^{k},B^{k})=\begin{cases}
\qquad\qquad\qquad\qquad\qquad\qquad\qquad0 & {\tt if}\quad\Vert A^{k}\Vert\Vert B^{k}\Vert<\tau\\[0.3cm]
\qquad\qquad\qquad\qquad\qquad\qquad\;\; A^{k}\cdot B^{k} & {\tt elseif}\quad k=k_{\textrm{max}}\\[0.3cm]
\left(\begin{array}{rr}
{\tt SpAMM}(A_{11}^{k+1},B_{11}^{k+1})\quad & {\tt SpAMM}(A_{11}^{k+1},B_{12}^{k+1})\\
+{\tt SpAMM}(A_{12}^{k+1},B_{21}^{k+1})\quad & +{\tt SpAMM}(A_{12}^{k+1},B_{22}^{k+1})\\[0.3cm]
{\tt SpAMM}(A_{21}^{k+1},B_{11}^{k+1})\quad & {\tt SpAMM}(A_{21}^{k+1},B_{12}^{k+1})\\
+{\tt SpAMM}(A_{22}^{k+1},B_{21}^{k+1})\quad & +{\tt SpAMM}(A_{22}^{k+1},B_{22}^{k+1})\end{array}\right) & {\tt else}\end{cases}\label{SpAMM}\end{equation}
 \end{widetext}The truncated product space accessed by SpAMM is shown
in Figure \ref{fig:Hierarchical-truncation-of} for matrices with
exponential and algebraic decay. 

At each tier in SpAMM, the local truncation error is bounded by $\Vert\widetilde{C}^{k}-C^{k}\Vert<\tau$,
with an error accumulation structurally similar to rounding in conventional
recursive multiplication, except that truncation is not guaranteed
to occur at each tier in the recursion, and truncation does not retain
even the approximate magnitude of the avoided sub-product. To see
the difference between truncation and round-off, consider the generic,
norm-wise bound for recursive multiplication employed by Demmel and
co-workers in Reference \cite{Demmel:2007:FastMM}: 

\begin{eqnarray}
\Vert\widetilde{C}-C\Vert & <\mu(n) & \epsilon\,\Vert A\Vert\Vert B\Vert\,+\mathcal{O}\left(\epsilon^{2}\right),\label{eq:RoundBound}\end{eqnarray}
with $\mu(n)$ a low order polynomial $\sim n^{d}$ and $d\geq1$.
Unlike rounding which is an error of commission, SpAMM creates errors
of omission. If $\tau<\Vert A\Vert\Vert B\Vert$, then no work is
performed and we obviously have$\Vert\widetilde{C}-C\Vert<\tau$,
which is different than $\Vert\widetilde{C}-C\Vert<\epsilon\Vert A\Vert\Vert B\Vert$
corresponding to the case of comparable round-off and truncation parameters
$\epsilon\sim\tau$. One could certainly bound SpAMM rigorously by
decreasing $\tau$ with increasing depth, $\tau^{k+1}=\tau^{k}/8$,
but that would be overly pessimistic, not taking into account signed
error accumulation or the localization and attenuation of errors due
to decay.

SpAMM is similar to the $\mathcal{H}$-matrix algebra of Hackbusch
and co-workers, where off-diagonal sub-matrices are treated as reduced
rank factorizations (truncated SVD), typically structured and grouped
to reflect properties of the underlying operators \cite{Grasedyck:2003:HMatrix}.
For problems with rapid decay, truncated SVD may behave in a similar
way to simple dropping schemes. SpAMM is different than the $\mathcal{H}$-matrix
algebra as it achieves separation uniquely in the product space and
does not rely on a reduced complexity representation of matrices.
For very slow decay, the $\mathcal{H}$-matrix algebra may certainly
offer a path for intractable problems for which SpAMM is ineffective. 

\begin{figure}[H]
\includegraphics[width=3.3in]{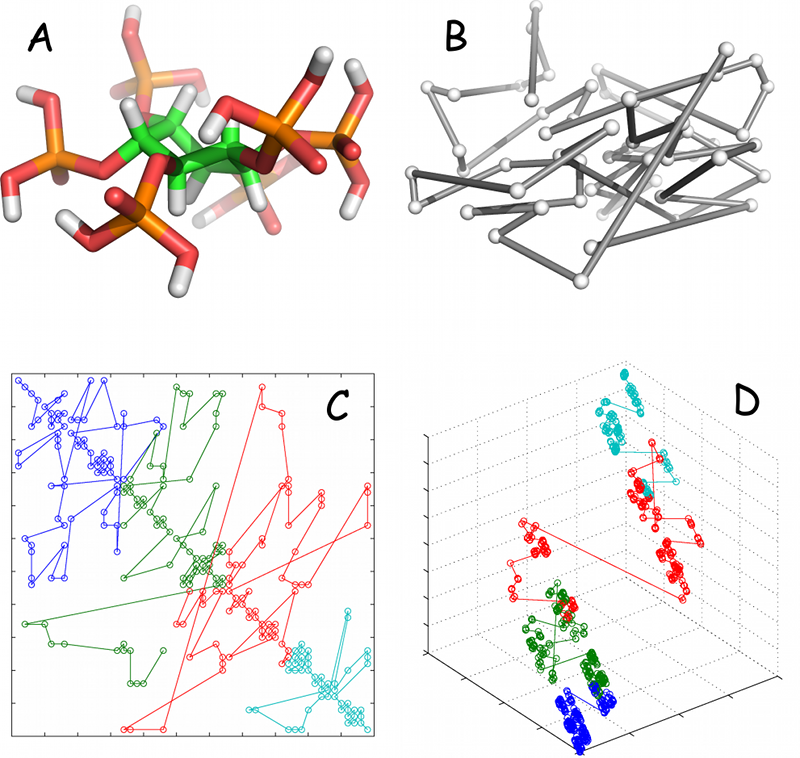}\caption{\label{fig:Exploiting-locality-Space}Space filling curves map atoms
in Cartesian space (A) onto an index that is locality preserving (B),
leading to clustering of matrix elements with respect to indicial
separation (C) when interactions are short ranged. The octree generated
by SpAMM in the three-dimensional product space, shown in Figure \ref{fig:Hierarchical-truncation-of},
is equivalent to a second space filling curve (spatial hashing) \cite{Warren:1992:HOT,Warren:1995:HOT,Samet:2006:DBDS}
that orders both matrices (C) and the product space (D), features
that can be exploited to achieve domain decomposition and load balance
(colors in C and D).}

\end{figure}

An understanding of the error accumulation in SpAMM must account for
the decay properties of $A$ and $B$, which in non-synthetic cases
are intimately related to the effects of ordering and structure of
the underlying physical, chemical or engineering application. Also,
ordering will determine the relative efficiencies of both matrix truncation
and SpAMM, particularly under blocking. The ordering used here is
based on the Space Filling Curve (SFC) method described in \cite{Challacombe:2000:SpMM}
(Hilbert atom ordering), and shown in Figure~\ref{fig:Exploiting-locality-Space},
involving also a second tier of ordering in the product space. The
first ordering maps atoms that are close in Cartesian-space to entries
close in the index space of the matrix. The second ordering is a natural
consequence of the SpAMM multiply, which recursively maps out a multi-level
octree (see also Figure~\ref{fig:Hierarchical-truncation-of}) with
cuboid coordinates that are equivalent to a spatial-hash; sorting
the hash produces a three-dimensional SFC in the product space \cite{Warren:1992:HOT,Warren:1995:HOT,Samet:2006:DBDS}.
This two-tiered structure is novel, providing an encompassing scheme
that parlays the locality of physical interactions into the data locality
of matrices (element clustering), and into the irregular product space.
This approach is applicable to other problems with underlying decay
properties, in which the finite-elements, radial basis functions,
\emph{etc.} replace atoms, or graph theoretical methods (nested dissection,
Cuthill-McKee, \emph{etc}.) replace the first-tier SFC altogether.
In either case, the correspondence between the recursive SpAMM product
space and a three-dimensional SFC provides a tool to exploit both
spatial and temporal locality in the distribution of work and data.

\section{Results and Discussion}

\begin{figure}[h]
\includegraphics[width=3.3in]{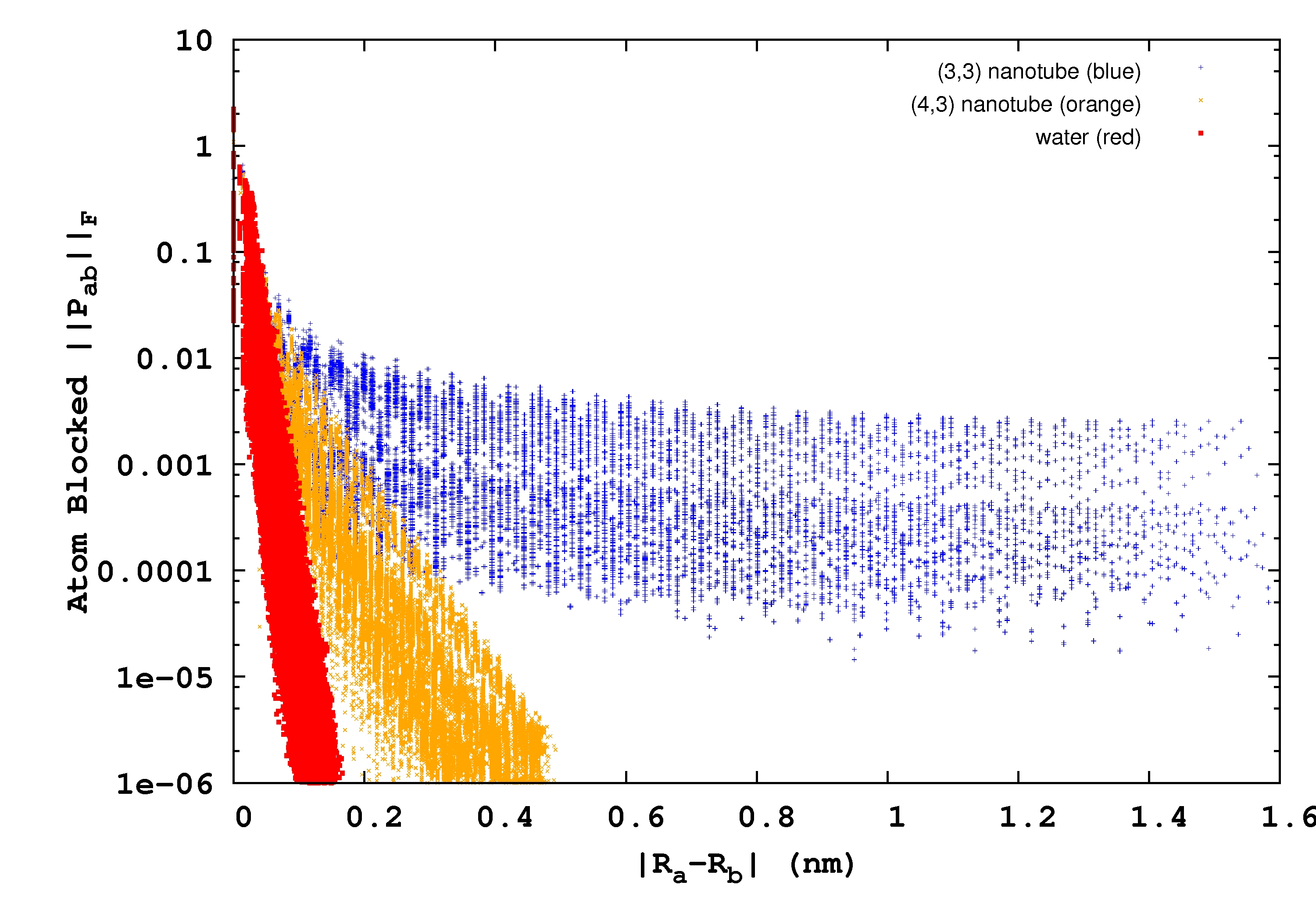}

\caption{Decay of normed density matrix atom-blocks $\left\Vert P_{ab}\right\Vert _{F}$
with Cartesian separation $\vec{|R_{a}}-\vec{R}_{b}|$ for the largest
molecules in each sequence: a 450 atom water cluster (red), a 752
atom (4,3) nanotube (orange) and a 780 atom (3,3) nanotube (blue).\label{fig:Decay-of-normed}}

\end{figure}

In this initial contribution, we limit ourselves to exploring the
numerical and computational behavior of SpAMM applied to problems
in electronic structure theory, and compare its relative merits to
the dropping of matrix elements. In ${\cal O}\left(n\right)$ electronic
structure calculations \cite{Goedecker:1999:ONREV,Goedecker:2003:ONREV,Li:2005:ONREV},
a primary source of error due to sparse matrix-multiplication develops
from early steps in the iterative construction of the Heaviside matrix
function $P=\theta[F-\mu I]$ (density matrix purification), which
is a projector of the effective Hamiltonian (Fockian) $F$ \cite{Niklasson:2003:TRS4}.
Starting from the basis spanning $F$, purification drives eigenvalues
to $0$s or $1$s, whereupon error accumulation due to an approximate
linear algebra is quenched \cite{Niklasson:2003:TRS4}. Under a given
approximate linear algebra, the electronic energy $E_{{\rm el}}={\rm Tr}(P.F)$
is a global measure of accumulated error. In the following Section,
we carry out purification on three molecular sequences of increasing
size, water clusters, (4,3) nanotubes and (3,3) nanotubes. In each
case, a Fockian $F$ was obtained from a fully converged Self-Consistent-Field
(SCF) cycle and used as basis for the purification; our numerical
experiments probe only errors within one density matrix solve (40-60
multiplies), and do not address error propagation throughout the SCF
cycle. As the rate of decay slows towards SCF convergence, these calculations
represent the instance of minimum decay, shown for each of the largest
molecular species in Figure~\ref{fig:Decay-of-normed}. Within each
sequence, the same number of iterative steps (40-60 matrix multiplications)
are taken using the TC2 purification algorithm \cite{Niklasson:2002:Pure}
and either: (I) matrix element dropping and exact multiplication or
(II) SpAMM. 

\begin{figure}[h]
\includegraphics[width=3.3in]{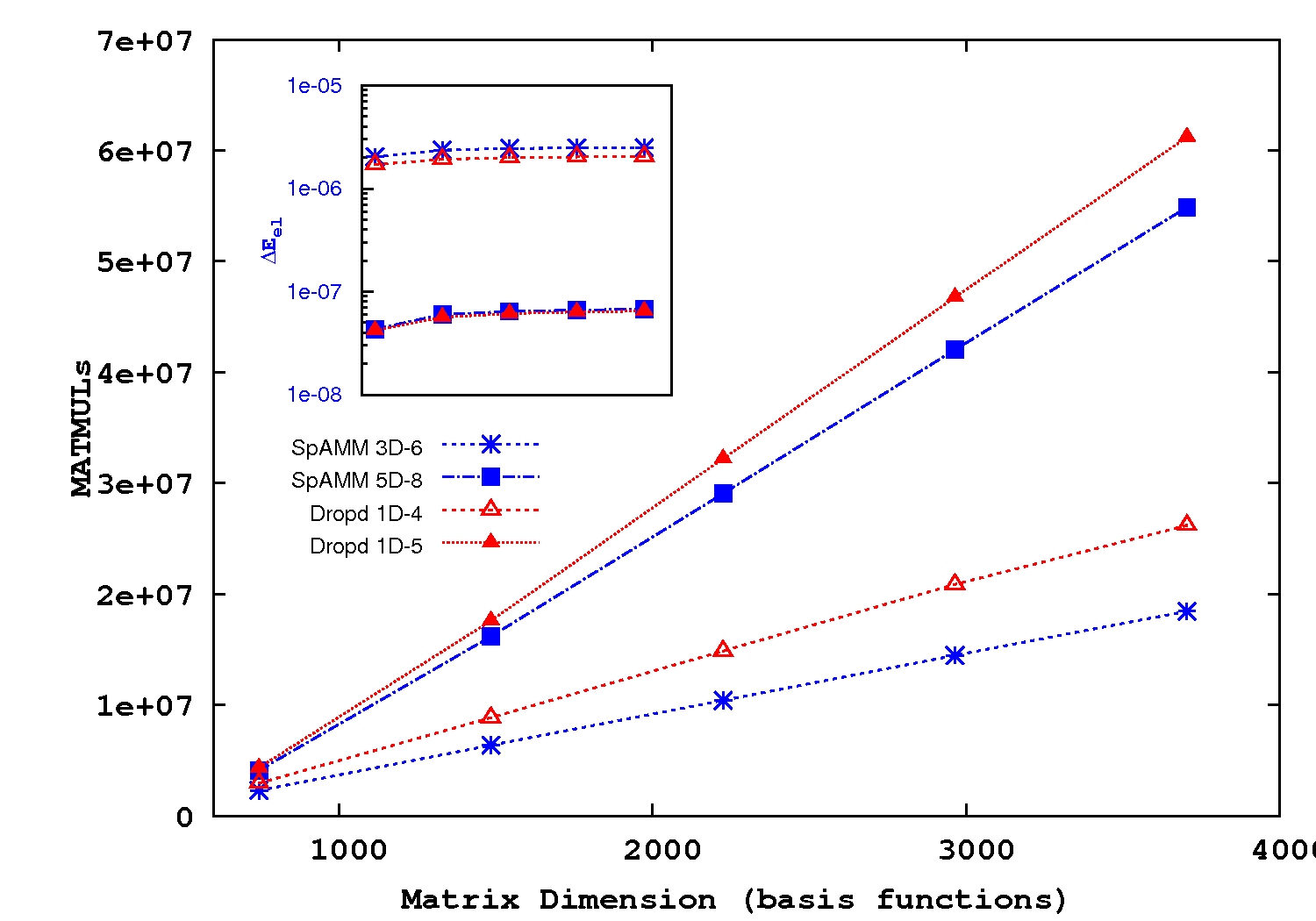}

\caption{Average number of 4x4 matrix multiplies (MATMULs) for a sequence of
(4,3) nanotubes at the RHF/STO-2G level of theory with dropping (red)
and SpAMM (blue).\label{fig:(4,3)}}

\end{figure}
\begin{figure}[h]
\includegraphics[width=3.3in]{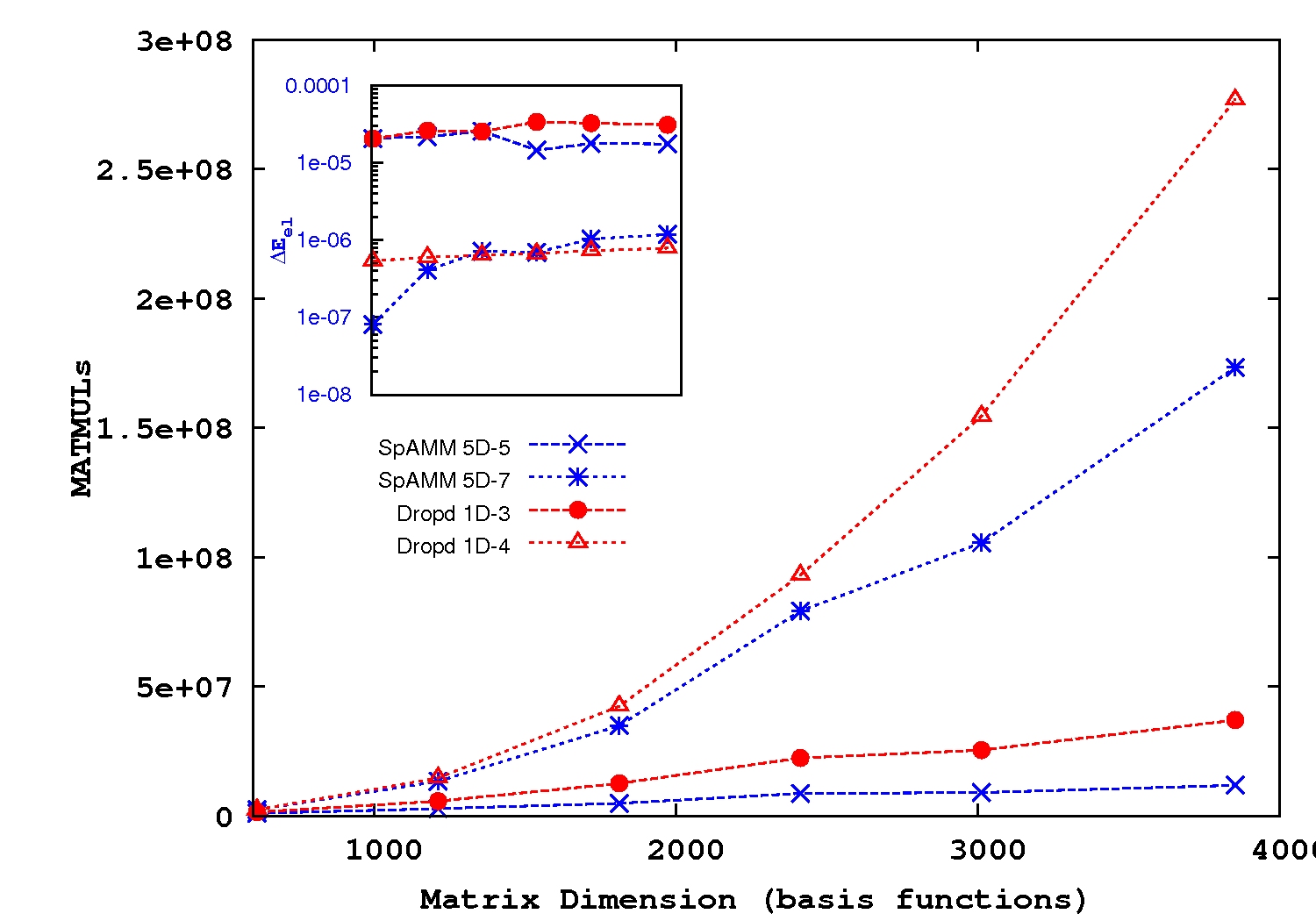}\caption{Average number of 4x4 matrix multiplies (MATMULs) for a sequence of
(3,3) nanotubes at the LDA/STO-2G/ level of theory with dropping (red)
and SpAMM (blue).\label{fig:(3,3)}}

\end{figure}
\begin{figure}[h]
\includegraphics[width=3.3in]{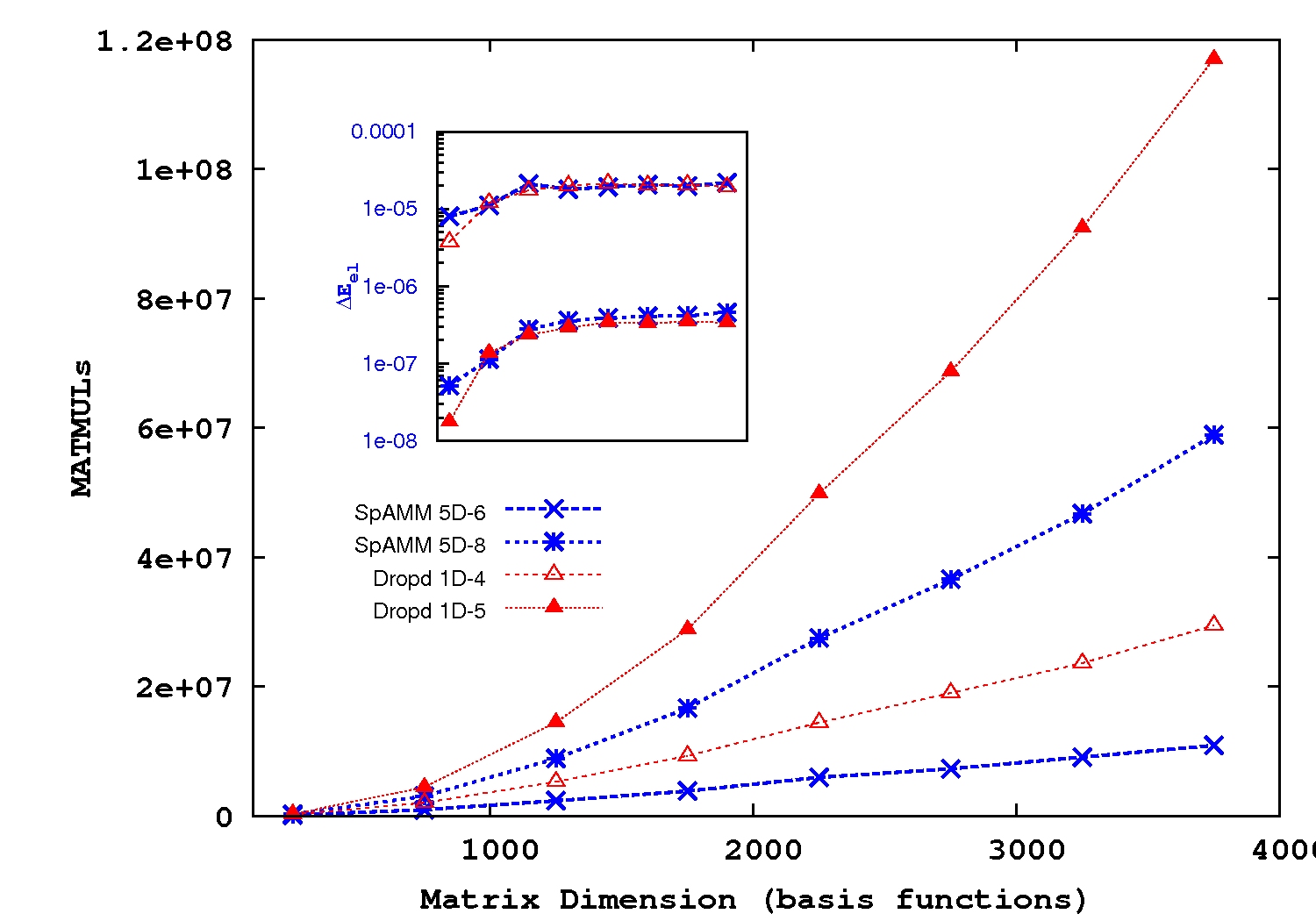}

\caption{Average number of 4x4 matrix multiplies (MATMULs) for a sequence of
water clusters at the RHF/6-31G{*}{*} level of theory with dropping
(red) and SpAMM (blue).\label{fig:Water}}

\end{figure}

In this study, we chose $k_{{\rm max}}$ to yield 4x4 blocks at the
finest level of resolution, corresponding to the most aggressive use
of single precision SSE on the x86 architecture. In all cases the
matrix norm employed is the Frobenius norm $\left\Vert \cdot\right\Vert \equiv\left\Vert \cdot\right\Vert _{F}$.
Thresholds, $\tau$, have been adjusted to roughly match relative
errors in the electronic energy,$\Delta E_{\textrm{el}}=|\tilde{E}_{{\rm el}}-E_{{\rm el}}|/|E_{{\rm el}}|$
between the two schemes, \noun{(I)} and (II), and the average number
of 4x4 MATMULs per purification step are reported. Multiplications
are only a proxy for CPU time, as neither case accounts for symbolic
overheads associated with the multiply (CSR, recursive-tree \emph{etc}.).
In the case of dropping, SpAMM was also used in the matrix multiply
but with zero threshold. After each multiplication in the dropping
scheme, a filter was applied to the resultant, dropping blocks at
the 4x4 level of resolution using the criteria $\left\Vert P^{k}\right\Vert _{F}<\tau$.
Results for the three molecular sequences are shown in Figures~\ref{fig:Water},
\ref{fig:(4,3)} and \ref{fig:(3,3)}.

For comparable values of $\Delta E_{\textrm{el}}$, SpAMM was found
to employ from slightly fewer multiplies for the (4,3) nanotube, to
dramatically less in the case of the water clusters. This result is
somewhat surprising, since the spatial decay is slower in the case
of the (4,3) nanotube than for the water clusters, as shown in Figure~\ref{fig:Decay-of-normed}.
While it seems reasonable to attribute this unexpected result to the
effects of dimensionality, further study is required to be sure. Its
also worth noting that the advantage of SpAMM relative to dropping
is brought down with decreasing $\tau$; for $\tau=0$ they both revert
to the same $\mathcal{O}(n^{3})$ complexity. For both systems with
exponential decay, error appears tightly controlled albeit within
an as yet unknown bound. Note however, that unlike matrix truncation
which leads to a product error that is $\mathcal{O}\left(\tau^{2}\right)$,
SpAMM leads to a truncation error that is $\mathcal{O}\left(\tau\right)$.

Comparing the quasi one-dimensional metallic (3,3) nanotube with slow
algebraic decay to the insulating (4,3) nanotube with exponential
decay, SpAMM gains substantially over dropping in the case of slower
decay. However, the ability of SpAMM to achieve a linear scaling complexity
in the case of the metallic system remains in question, as in the
case of the tightest threshold, the SpAMM errors do not appear to
be well controlled, and the cost does not appear linear, at least
not in this size regime. On the other hand, the SpAMM result does
enjoy a significant reduction in cost relative to dropping, and the
error increase is modest.

\section{Conclusion\label{sec:Conclusion}}

The Sparse Approximate Matrix Multiply (SpAMM) is a fast method for
matrices with decay, which is different from element dropping or the
$\mathcal{H}$-algebra in that it uniquely involves truncation of
the product space rather than the matrix space. For matrices with
exponential or fast algebraic decay, SpAMM can achieve stable error
control comparable to element dropping, but with a greatly reduced
number of floating point operations. The results presented here are
preliminary, and have not yet explored the interesting problems of
slow decay in the asymptotic limit, ordering, error bounds, the cost
of recursion, high performance implementations or broader gauges of
accuracy and efficiency, such as the use of SpAMM in the context of
the Self-Consistent-Field cycle. Of particular concern is the relationship
between complexity, matrix decay and error control. Based on our numerical
tests, we postulate that the algorithm is at worst $\mathcal{O}(n\lg n)$
for matrices with sufficiently fast decay. 

In addition to similarities with the $\mathcal{H}$-algebra, SpAMM
falls under the rubric of the generalized $N$-body problem \cite{Gray:2001:NBody}.
From this perspective, its worth noting also the connection between
matrix-matrix multiplication as $N$-body problem and matrix-matrix
multiplication as spatial join \cite{Amossen:2009:SpMMeqJoin}, as
well as between $N$-body problems and data base theory in general
\cite{Samet:2006:DBDS}. 

Next, we draw attention to the second tier of Space Filling Curve
(SFC) shown in Figure~\ref{fig:Exploiting-locality-Space}, which
provides a mechanism for domain decomposition and load balance that
is proven for parallel irregular problems \cite{Warren:1992:HOT,Warren:1995:HOT,Aluru:1997:SFC,Devine:2005:SFC}.
Also, the improvement gained in Reference \cite{Buloc:2008:SpMM}
on going from a one-dimensional to a two-dimensional matrix partitioning
scheme for the parallel SpMM suggests that partitioning the three-dimensional
product space instead may provide an even higher degree of flexibility
and granularity.

The authors acknowledge support through Los Alamos LDRD award ER20110230
(computational co-design) as well as funds from the U.S. Department
of Energy. Los Alamos National Laboratory is operated by the Los Alamos
National Security, LLC, for the National Nuclear Security Administration
of the U.S. Department of Energy under Contract No. DE-AC52-06NA25396.
Special acknowledgments go to the International Ten Bar Café for tasty
libations in a scientific and collegial atmosphere, and to Michele
Benzi for valuable input.

\bibliography{SpAMM}

\begin{thebibliography}{29}
\expandafter\ifx\csname natexlab\endcsname\relax\def\natexlab#1{#1}\fi
\expandafter\ifx\csname bibnamefont\endcsname\relax
  \def\bibnamefont#1{#1}\fi
\expandafter\ifx\csname bibfnamefont\endcsname\relax
  \def\bibfnamefont#1{#1}\fi
\expandafter\ifx\csname citenamefont\endcsname\relax
  \def\citenamefont#1{#1}\fi
\expandafter\ifx\csname url\endcsname\relax
  \def\url#1{\texttt{#1}}\fi
\expandafter\ifx\csname urlprefix\endcsname\relax\def\urlprefix{URL }\fi
\providecommand{\bibinfo}[2]{#2}
\providecommand{\eprint}[2][]{\url{#2}}

\bibitem[{\citenamefont{Demmel and Higham}(1992)}]{Demmel:1992:FastMM}
\bibinfo{author}{\bibfnamefont{J.~W.} \bibnamefont{Demmel}} \bibnamefont{and}
  \bibinfo{author}{\bibfnamefont{N.~J.} \bibnamefont{Higham}},
  \bibinfo{journal}{ACM Trans. Math. Softw.} \textbf{\bibinfo{volume}{18}},
  \bibinfo{pages}{274} (\bibinfo{year}{1992}).

\bibitem[{\citenamefont{Demmel et~al.}(2007)\citenamefont{Demmel, Dumitriu,
  Holtz, and Kleinberg}}]{Demmel:2007:FastMM}
\bibinfo{author}{\bibfnamefont{J.}~\bibnamefont{Demmel}},
  \bibinfo{author}{\bibfnamefont{I.}~\bibnamefont{Dumitriu}},
  \bibinfo{author}{\bibfnamefont{O.}~\bibnamefont{Holtz}}, \bibnamefont{and}
  \bibinfo{author}{\bibfnamefont{R.}~\bibnamefont{Kleinberg}},
  \bibinfo{journal}{Numer. Math.} \textbf{\bibinfo{volume}{106}},
  \bibinfo{pages}{199} (\bibinfo{year}{2007}).

\bibitem[{\citenamefont{Yuster and Zwick}(2005)}]{Yuster:2005:FastMM}
\bibinfo{author}{\bibfnamefont{R.}~\bibnamefont{Yuster}} \bibnamefont{and}
  \bibinfo{author}{\bibfnamefont{U.}~\bibnamefont{Zwick}},
  \bibinfo{journal}{ACM Transactions on Algorithms}
  \textbf{\bibinfo{volume}{1}}, \bibinfo{pages}{2} (\bibinfo{year}{2005}).

\bibitem[{\citenamefont{Benzi and Razouk}(2007)}]{Benzi:2007:Decay}
\bibinfo{author}{\bibfnamefont{M.}~\bibnamefont{Benzi}} \bibnamefont{and}
  \bibinfo{author}{\bibfnamefont{N.}~\bibnamefont{Razouk}},
  \bibinfo{journal}{Electronic Transactions on Numerical Analysis}
  \textbf{\bibinfo{volume}{28}}, \bibinfo{pages}{16} (\bibinfo{year}{2007}).

\bibitem[{\citenamefont{Benzi et~al.}(2010)\citenamefont{Benzi, Boito, and
  Razouk}}]{Benzi:2010:Decay}
\bibinfo{author}{\bibfnamefont{M.}~\bibnamefont{Benzi}},
  \bibinfo{author}{\bibfnamefont{P.}~\bibnamefont{Boito}}, \bibnamefont{and}
  \bibinfo{author}{\bibfnamefont{N.}~\bibnamefont{Razouk}},
  \bibinfo{journal}{Manuscript in preparation}  (\bibinfo{year}{2010}).

\bibitem[{\citenamefont{Benzi and Tuma}(2000)}]{Benzi:2000:Inv}
\bibinfo{author}{\bibfnamefont{M.}~\bibnamefont{Benzi}} \bibnamefont{and}
  \bibinfo{author}{\bibfnamefont{M.}~\bibnamefont{Tuma}},
  \bibinfo{journal}{SIAM Journal on Scientific Computing}
  \textbf{\bibinfo{volume}{21}}, \bibinfo{pages}{1851} (\bibinfo{year}{2000}).

\bibitem[{\citenamefont{Iserles}(1999)}]{Iserles:1999:Decay}
\bibinfo{author}{\bibfnamefont{A.}~\bibnamefont{Iserles}},
  \emph{\bibinfo{title}{How large is the exponential of a banded matrix?}}
  (\bibinfo{year}{1999}).

\bibitem[{\citenamefont{Challacombe}(1999)}]{Challacombe:1999:DMM}
\bibinfo{author}{\bibfnamefont{M.}~\bibnamefont{Challacombe}},
  \bibinfo{journal}{J. Chem. Phys.} \textbf{\bibinfo{volume}{110}},
  \bibinfo{pages}{2332} (\bibinfo{year}{1999}).

\bibitem[{\citenamefont{Challacombe}(2000)}]{Challacombe:2000:SpMM}
\bibinfo{author}{\bibfnamefont{M.}~\bibnamefont{Challacombe}},
  \bibinfo{journal}{Compututer Physics Communications}
  \textbf{\bibinfo{volume}{128}}, \bibinfo{pages}{93} (\bibinfo{year}{2000}).

\bibitem[{\citenamefont{Simoncini and Elden}(2002)}]{Simoncini:2002:Inexact}
\bibinfo{author}{\bibfnamefont{V.}~\bibnamefont{Simoncini}} \bibnamefont{and}
  \bibinfo{author}{\bibfnamefont{L.}~\bibnamefont{Elden}},
  \bibinfo{journal}{Bit Numerical Mathematics} \textbf{\bibinfo{volume}{42}},
  \bibinfo{pages}{159} (\bibinfo{year}{2002}).

\bibitem[{\citenamefont{Simoncini and Szyld}(2003)}]{Simoncini:2003:Inexact}
\bibinfo{author}{\bibfnamefont{V.}~\bibnamefont{Simoncini}} \bibnamefont{and}
  \bibinfo{author}{\bibfnamefont{D.~B.} \bibnamefont{Szyld}},
  \bibinfo{journal}{SIAM Journal on Scientific Computing}
  \textbf{\bibinfo{volume}{25}}, \bibinfo{pages}{454} (\bibinfo{year}{2003}).

\bibitem[{\citenamefont{Simoncini and Szyld}(2005)}]{Simoncini:2005:Inexact}
\bibinfo{author}{\bibfnamefont{V.}~\bibnamefont{Simoncini}} \bibnamefont{and}
  \bibinfo{author}{\bibfnamefont{D.~B.} \bibnamefont{Szyld}},
  \bibinfo{journal}{SIAM Review} \textbf{\bibinfo{volume}{47}},
  \bibinfo{pages}{247} (\bibinfo{year}{2005}).

\bibitem[{\citenamefont{Challacombe}(2010)}]{Challacombe:2010:TDSCF}
\bibinfo{author}{\bibfnamefont{M.}~\bibnamefont{Challacombe}},
  \bibinfo{journal}{arXiv} \textbf{\bibinfo{volume}{quant-ph}},
  \bibinfo{pages}{1001.2586} (\bibinfo{year}{2010}).

\bibitem[{\citenamefont{Galli}(1996)}]{Galli:1996:ONREV}
\bibinfo{author}{\bibfnamefont{G.}~\bibnamefont{Galli}},
  \bibinfo{journal}{Current Opinion in Solid State \& Materials Science}
  \textbf{\bibinfo{volume}{1}}, \bibinfo{pages}{864} (\bibinfo{year}{1996}).

\bibitem[{\citenamefont{Goedecker}(1999)}]{Goedecker:1999:ONREV}
\bibinfo{author}{\bibfnamefont{S.}~\bibnamefont{Goedecker}},
  \bibinfo{journal}{Reviews of Modern Physics} \textbf{\bibinfo{volume}{71}},
  \bibinfo{pages}{1085} (\bibinfo{year}{1999}).

\bibitem[{\citenamefont{Goedecker and Scuseria}(2003)}]{Goedecker:2003:ONREV}
\bibinfo{author}{\bibfnamefont{S.}~\bibnamefont{Goedecker}} \bibnamefont{and}
  \bibinfo{author}{\bibfnamefont{G.}~\bibnamefont{Scuseria}},
  \bibinfo{journal}{Computing in Science Engineering}
  \textbf{\bibinfo{volume}{5}}, \bibinfo{pages}{14 } (\bibinfo{year}{2003}).

\bibitem[{\citenamefont{Li et~al.}(2005)\citenamefont{Li, He, and
  Yang}}]{Li:2005:ONREV}
\bibinfo{author}{\bibfnamefont{Z.~Y.} \bibnamefont{Li}},
  \bibinfo{author}{\bibfnamefont{W.}~\bibnamefont{He}}, \bibnamefont{and}
  \bibinfo{author}{\bibfnamefont{J.~L.} \bibnamefont{Yang}},
  \bibinfo{journal}{Progress in Chemistry} \textbf{\bibinfo{volume}{17}},
  \bibinfo{pages}{192} (\bibinfo{year}{2005}).

\bibitem[{\citenamefont{Gray and Moore}(2001)}]{Gray:2001:NBody}
\bibinfo{author}{\bibfnamefont{A.~G.} \bibnamefont{Gray}} \bibnamefont{and}
  \bibinfo{author}{\bibfnamefont{A.~W.} \bibnamefont{Moore}}, in
  \emph{\bibinfo{booktitle}{Advances in Neural Information Processing Systems}}
  (\bibinfo{publisher}{MIT Press}, \bibinfo{year}{2001}),
  vol.~\bibinfo{volume}{4}, pp. \bibinfo{pages}{521--527}.

\bibitem[{\citenamefont{Bini and Lotti}(1980)}]{Bini:1980:FastMM}
\bibinfo{author}{\bibfnamefont{D.}~\bibnamefont{Bini}} \bibnamefont{and}
  \bibinfo{author}{\bibfnamefont{G.}~\bibnamefont{Lotti}},
  \bibinfo{journal}{Numerische Mathematik} \textbf{\bibinfo{volume}{36}},
  \bibinfo{pages}{63} (\bibinfo{year}{1980}).

\bibitem[{\citenamefont{Grasedyck and
  Hackbusch}(2003)}]{Grasedyck:2003:HMatrix}
\bibinfo{author}{\bibfnamefont{L.}~\bibnamefont{Grasedyck}} \bibnamefont{and}
  \bibinfo{author}{\bibfnamefont{W.}~\bibnamefont{Hackbusch}},
  \bibinfo{journal}{Computing} \textbf{\bibinfo{volume}{70}},
  \bibinfo{pages}{2003} (\bibinfo{year}{2003}).

\bibitem[{\citenamefont{Warren and Salmon}(1992)}]{Warren:1992:HOT}
\bibinfo{author}{\bibfnamefont{M.~S.} \bibnamefont{Warren}} \bibnamefont{and}
  \bibinfo{author}{\bibfnamefont{J.~K.} \bibnamefont{Salmon}}, in
  \emph{\bibinfo{booktitle}{Supercomputing '92}} (\bibinfo{publisher}{IEEE
  Comp. Soc.}, \bibinfo{address}{Los Alamitos}, \bibinfo{year}{1992}), pp.
  \bibinfo{pages}{570--576}, \bibinfo{note}{(1992 Gordon Bell Prize winner)}.

\bibitem[{\citenamefont{Warren and Salmon}(1995)}]{Warren:1995:HOT}
\bibinfo{author}{\bibfnamefont{M.~S.} \bibnamefont{Warren}} \bibnamefont{and}
  \bibinfo{author}{\bibfnamefont{J.~K.} \bibnamefont{Salmon}},
  \emph{\bibinfo{title}{A parallel, portable and versatile treecode}}
  (\bibinfo{publisher}{SIAM}, \bibinfo{address}{Philadelphia},
  \bibinfo{year}{1995}), chap.~\bibinfo{chapter}{1}.

\bibitem[{\citenamefont{Samet}(2006)}]{Samet:2006:DBDS}
\bibinfo{author}{\bibfnamefont{H.}~\bibnamefont{Samet}},
  \emph{\bibinfo{title}{{Foundations of Multidimensional and Metric Data
  Structures}}} (\bibinfo{publisher}{Morgan Kaufmann}, \bibinfo{year}{2006}).

\bibitem[{\citenamefont{Niklasson et~al.}(2003)\citenamefont{Niklasson,
  Tymczak, and Challacombe}}]{Niklasson:2003:TRS4}
\bibinfo{author}{\bibfnamefont{A.~M.~N.} \bibnamefont{Niklasson}},
  \bibinfo{author}{\bibfnamefont{C.~J.} \bibnamefont{Tymczak}},
  \bibnamefont{and}
  \bibinfo{author}{\bibfnamefont{M.}~\bibnamefont{Challacombe}},
  \bibinfo{journal}{J. Comp. Phys.} \textbf{\bibinfo{volume}{118}},
  \bibinfo{pages}{8611} (\bibinfo{year}{2003}).

\bibitem[{\citenamefont{Niklasson}(2002)}]{Niklasson:2002:Pure}
\bibinfo{author}{\bibfnamefont{A.~M.~N.} \bibnamefont{Niklasson}},
  \bibinfo{journal}{Physical Review B} \textbf{\bibinfo{volume}{66}},
  \bibinfo{pages}{5} (\bibinfo{year}{2002}).

\bibitem[{\citenamefont{Amossen and Pagh}(2009)}]{Amossen:2009:SpMMeqJoin}
\bibinfo{author}{\bibfnamefont{R.}~\bibnamefont{Amossen}} \bibnamefont{and}
  \bibinfo{author}{\bibfnamefont{R.}~\bibnamefont{Pagh}}, in
  \emph{\bibinfo{booktitle}{Proceedings of the 12th International Conference on
  Database Theory}} (\bibinfo{publisher}{ACM}, \bibinfo{year}{2009}), pp.
  \bibinfo{pages}{121--126}.

\bibitem[{\citenamefont{Aluru and Sevilgen}(1997)}]{Aluru:1997:SFC}
\bibinfo{author}{\bibfnamefont{S.}~\bibnamefont{Aluru}} \bibnamefont{and}
  \bibinfo{author}{\bibfnamefont{F.~E.} \bibnamefont{Sevilgen}}, in
  \emph{\bibinfo{booktitle}{Proceedings of the 4th {IEEE} Conference on High
  Performance Computing}} (\bibinfo{year}{1997}), pp.
  \bibinfo{pages}{230--235}.

\bibitem[{\citenamefont{Devine et~al.}(2005)\citenamefont{Devine, Boman,
  Heaphy, Hendrickson, Teresco, Faik, Flaherty, and
  Gervasio}}]{Devine:2005:SFC}
\bibinfo{author}{\bibfnamefont{K.~D.} \bibnamefont{Devine}},
  \bibinfo{author}{\bibfnamefont{E.~G.} \bibnamefont{Boman}},
  \bibinfo{author}{\bibfnamefont{R.~T.} \bibnamefont{Heaphy}},
  \bibinfo{author}{\bibfnamefont{B.~A.} \bibnamefont{Hendrickson}},
  \bibinfo{author}{\bibfnamefont{J.~D.} \bibnamefont{Teresco}},
  \bibinfo{author}{\bibfnamefont{J.}~\bibnamefont{Faik}},
  \bibinfo{author}{\bibfnamefont{J.~E.} \bibnamefont{Flaherty}},
  \bibnamefont{and} \bibinfo{author}{\bibfnamefont{L.~G.}
  \bibnamefont{Gervasio}}, \bibinfo{journal}{Applied Numerical Mathematics}
  \textbf{\bibinfo{volume}{52}}, \bibinfo{pages}{133 } (\bibinfo{year}{2005}),
  \bibinfo{note}{{ADAPT} '03: Conference on Adaptive Methods for Partial
  Differential Equations and Large-Scale Computation}.

\bibitem[{\citenamefont{Buluc and Gilbert}(2008)}]{Buloc:2008:SpMM}
\bibinfo{author}{\bibfnamefont{A.}~\bibnamefont{Buluc}} \bibnamefont{and}
  \bibinfo{author}{\bibfnamefont{J.~R.} \bibnamefont{Gilbert}}, in
  \emph{\bibinfo{booktitle}{{ICPP} '08: Proceedings of the 2008 37th
  International Conference on Parallel Processing}} (\bibinfo{publisher}{IEEE
  Computer Society}, \bibinfo{address}{Washington, DC, USA},
  \bibinfo{year}{2008}), pp. \bibinfo{pages}{503--510}.

\end{thebibliography}

\end{document}